# Beyond Bonferroni: Hierarchical Multiple Testing in Empirical Research

Sebastian Calónico
*UC Davis*

Sebastian Galiani
*University of Maryland*
and *NBER*

July 2025

Abstract

Empirical research in the social and medical sciences frequently involves testing multiple hypotheses simultaneously, increasing the risk of false positives due to chance. Classical multiple testing procedures, such as the Bonferroni correction, control the family-wise error rate (FWER) but tend to be overly conservative, reducing statistical power. Stepwise alternatives like the Holm and Hochberg procedures offer improved power while maintaining error control under certain dependence structures. However, these standard approaches typically ignore hierarchical relationships among hypotheses—structures that are common in settings such as clinical trials and program evaluations, where outcomes are often logically or causally linked. Hierarchical multiple testing procedures—including fixed sequence, fallback, and gatekeeping methods—explicitly incorporate these relationships, providing more powerful and interpretable frameworks for inference. This paper reviews key hierarchical methods, compares their statistical properties and practical trade-offs, and discusses implications for applied empirical research.

## 1. Introduction

Empirical research in the social and medical sciences frequently involves testing multiple hypotheses simultaneously. Without appropriate adjustments, this practice inflates the probability of making at least one Type I error—that is, incorrectly rejecting a true null hypothesis—purely by chance. For example, if m independent hypotheses are tested at a significance level $\alpha = 0.05$, the family-wise error rate (FWER), or the probability of making at least one false rejection, is $1-(1-0.05)^{m} \geq 0.05$, which increases rapidly with the number of tests.

Classical multiple testing procedures are designed to control the FWER. The Bonferroni correction, one of the most widely used methods, adjusts the significance level of each test, ensuring strong control of the FWER regardless of the number or correlation of tests. However, it is often overly conservative, especially when m is large or when test statistics are positively correlated, resulting in a substantial loss of statistical power. Several alternatives have been proposed. Stepwise procedures (Holm, 1979; Hochberg, 1988) maintain control of the FWER under specific dependence assumptions while offering improved power relative to Bonferroni. These methods sequentially adjust significance thresholds based on the ordering of p-values, allowing greater flexibility in hypothesis testing. Resampling-based approaches (Westfall and Young, 1993) provide a powerful alternative by estimating the joint distribution of test statistics under the null through permutation or bootstrap methods. This allows for valid inference under complex dependence structures without relying on strong parametric assumptions.

While these procedures represent important advances in multiple testing methodology, they generally treat hypotheses as independent or exchangeable, ignoring potential logical or structural relationships among them. In many empirical applications, however, hypotheses follow a natural ordering or are logically or causally related. Such hierarchical structures frequently arise in clinical trials (e.g., primary and secondary endpoints), program evaluations (e.g., overall effects and subgroup analyses), and other multilevel designs. Standard multiple testing methods fail to exploit this structure, potentially missing opportunities for both improved power and interpretability. Hierarchical multiple testing procedures—such as fixed sequence, fallback, and gatekeeping methods—incorporate these dependencies by testing hypotheses in a prespecified order or under logical constraints. These approaches offer a more principled and efficient framework for inference in structured testing environments. This paper reviews and compares key hierarchical procedures, highlighting their theoretical properties, practical implementation, and relevance for applied research in the social and medical sciences.

## 2. Classical, Stepwise and Resampling Methods for FWER Control

The Bonferroni correction is the simplest and most widely used method for controlling the FWER. It adjusts the significance level for each individual hypothesis test to $\alpha/m$, where $\alpha$ is the desired overall error rate and m is the number of hypotheses. While it guarantees strong control of the FWER under any dependence structure, it is often overly conservative, especially when the number of tests is large.

The Holm step-down procedure improves upon Bonferroni by sequentially testing hypothesis ordered by their associated p-values. Let $p(1) \leq p(2) \leq \cdots \leq p(m)$ denote the ordered p-values. Starting with the smallest, each p(i) is compared to $\alpha/(m-i+1)$. Testing stops at the first non-rejection of the null hypothesis of no effect, and all subsequent null hypotheses are not rejected. Holm's method is uniformly more powerful than Bonferroni and maintains strong FWER control under arbitrary dependence. The Hochberg step-up procedure also uses ordered p-values but tests them in reverse order. Starting with the largest p(m), each p(i) is compared to $\alpha/(m-i+1)$. The largest index i such that $p(i) \leq \alpha/(m-i+1)$ is identified, and all hypotheses with $p(j) \leq p(i)$ for $j \leq i$ are rejected. Hochberg's method is more powerful than Holm's but requires the assumption that test statistics are independent or positively dependent.

Resampling approaches to multiple testing—such as permutation and bootstrap methods—offer an alternative framework for adjusting p-values while accounting for some dependence structure among test statistics. These methods are particularly useful when classical procedures (e.g., Bonferroni, Holm) are too conservative due to correlation among tests or when distributional assumptions are difficult to justify. Resampling generates a reference distribution for test statistics (or p-values) under the global null hypothesis by repeatedly re-analyzing the data after randomly permuting labels or resampling observations. These resamples reflect what one would expect by chance alone, thereby providing a data-driven way to assess the extremity of observed results.

Two common approaches include permutation tests (which reshuffle the data under the null hypothesis, maintaining the structure of the data), and bootstrap methods (where samples are drawn with replacement to approximate the sampling distribution of test statistics).

Unlike methods that assume independence, resampling methods preserve the correlation structure among tests without relying on strong distributional assumptions. Westfall and Young (1993) propose a free step-down resampling methodology for controlling the FWER using adjusted p-values that account for the joint distribution of test statistics. Unlike fixed corrections (e.g., Bonferroni), this method uses the empirical dependence structure in the data to provide sharper and less conservative adjustments, particularly under subset pivotality[1]. Suppose we are testing m null hypotheses $H_1,\ldots,H_m$ with corresponding test statistics $T_1,\ldots,T_m$. Let $p_1,\ldots,p_m$ be the unadjusted p-values.

1. Generate B resampled datasets under the complete null (e.g., via permutations or bootstrapping), preserving the dependence structure.
2. For each resample $b=1,\ldots,B$, compute the test statistics $T_{1(b)},\ldots,T_{m(b)}$ , and obtain their p-values
3. Step-down adjustment:
    - Order the observed p-values: $p_{(1)}\leq\cdots\leq p_{(m)}$.
    - For each hypothesis $H_{(i)}$ , define its adjusted p-value as the proportion of resamples in which the minimum p-value among all hypotheses not yet rejected is less than or equal to $p_{(i)}$.
4. Proceed stepwise: reject $H_{(1)}$ if its adjusted p-value $\leq \alpha$, then check $H_{(2)}$, and so on, stopping when a p-value exceeds $\alpha$.

By tracking the largest test statistic in each resample, the method constructs a null distribution that reflects the worst-case outcomes due to chance alone. A p-value is then adjusted based on how often this null distribution produces a statistic at least as extreme as the observed one.

Despite its strengths, the method has several limitations. It is computationally intensive, becoming demanding for large datasets or complex models. It also requires careful implementation:

[1] A set of test statistics satisfies subset pivotality if the joint distribution of the test statistics under any subset of true null hypotheses is the same as the distribution under the complete null (i.e., when all nulls are true). The joint distribution of p-values for the true null hypotheses does not depend on which specific ones are true. Without this condition, resampling from the complete null may lead to invalid inference, especially when there are many false nulls.

resampling procedures must preserve the null hypothesis, and the step-down variant depends on the subset pivotality assumption, which may not always hold. Moreover, while it effectively controls FWER, this can be overly conservative in settings with many hypotheses, where false discovery rate (FDR) control might be preferable.

## 3. Hierarchical testing of multiple hypotheses

In many applications, including both randomized experiments and observational studies, hypotheses are logically or causally related, forming a hierarchical structure. Standard multiple testing corrections often ignore these relationships, potentially leading to overly conservative conclusions. Hierarchical procedures explicitly account for such dependencies, improving interpretability and power. For instance, Cattaneo et al. (2009) evaluate Piso Firme, a large-scale Mexican program that replaced dirt floors with cement floors in low-income households. The authors assess multiple outcomes, including child health and adult well-being, many of which are causally linked. They find that the program significantly reduces parasitic infestations, diarrhea, and anemia among children, and improves cognitive development. These findings support a hypothesized causal chain: replacing dirt floors disrupts parasite transmission, leading to fewer infections and less anemia, which in turn improves cognitive outcomes. Such structured hypotheses lend themselves naturally to hierarchical testing, where the significance of downstream effects (e.g., cognitive development) can be interpreted more robustly when upstream mechanisms (e.g., parasite reduction) are established.

Hierarchical structures have important practical implications for empirical research. In the Piso Firme study, for example, the authors examine the impact of the program on more than 30 outcomes, raising concerns about controlling the FWER. Standard multiple testing procedures, while they may account for correlations among test statistics, typically ignore logical or causal relationships among hypotheses. As a result, they may reduce power and obscure meaningful effects. Hierarchical multiple testing procedures provide a more structured and interpretable alternative, particularly well-suited to settings like this one. We focus on three tractable approaches: fixed sequence, fallback and gatekeeping procedures.

**Fixed sequence procedures** are a class of hierarchical multiple testing methods in which hypotheses are tested in a pre-specified order, typically reflecting their logical, theoretical, or clinical importance. Each hypothesis $H_i$ is tested at the full significance level $\alpha$, without any adjustment for multiplicity. However, testing proceeds sequentially and stops as soon as a hypothesis fails to be rejected. This rule ensures strong control of the FWER under arbitrary dependence among test statistics. The approach is particularly powerful when early hypotheses in the sequence are likely to be true alternatives, as it avoids diluting the significance level across tests. Its simplicity and interpretability make it an appealing option in settings where a natural or policy-relevant ordering of hypotheses exists.

**Fallback procedures**, first introduced by Wiens (2003), offer a flexible sequential testing framework that improves power over standard Bonferroni-type corrections. Unlike fixed sequence procedures, which terminate testing upon the first non-rejection, the fallback method allows all hypotheses in a pre-specified sequence to be tested, regardless of earlier results.

The overall significance level $\alpha$ is partitioned among the k hypotheses according to predefined weights: each hypothesis $H_i$ is initially allocated $\alpha_i = w_i \times \alpha$, where $\sum w_i = 1$, analogous to a weighted Bonferroni correction. However, testing proceeds sequentially, and the local significance level can increase as testing progresses. Specifically, the hypothesis $H_i$ is tested at level:

$$\alpha_i' = \alpha_i \quad \text{if } H_{\{i-1\}} \text{ is not rejected}$$

$$= \alpha_i + \alpha'_{\{i-1\}} \quad \text{if } H_{\{i-1\}} \text{ is rejected}$$

This recursive redistribution of unused significance levels—often called "α propagation"—allows the procedure to maintain strong FWER control while increasing power relative to Bonferroni, particularly when early hypotheses are expected to show effects but do not reach statistical significance.

While the significance level for testing a hypothesis with the fallback depends on whether the previous hypothesis in the sequence was rejected, a more general approach is to allow the significance level for testing a hypothesis later in the sequence to depend on the level of evidence of testing earlier hypotheses. Huque and Alosh (2008) proposed hierarchical testing while accounting for correlation between hypotheses. Xie (2012) extends this model for when there are four or more endpoints using a weighted correction for correlated tests. An alternative approach, introduced by Li and Mehrotra (2008), is the adaptive alpha allocation.

**Gatekeeping procedures** are a class of hierarchical multiple testing methods designed to control the FWER across ordered families of hypotheses, reflecting logical, temporal, or clinical priorities. Originally developed in the context of regulatory trials (Dmitrienko et al., 2003; Westfall and Krishen, 2001), these procedures allow researchers to structure hypothesis testing so that confirmatory claims are made only when pre-specified criteria are met in earlier stages of testing.

Let $F_1, F_2, \ldots, F_K$ be K ordered families of hypotheses. Gatekeeping procedures allocate the overall significance level α across these families and specify rules for propagating testing rights. In serial gatekeeping, family $F_{\{k+1\}}$ is tested only if all hypotheses in $F_k$ are rejected, while in parallel gatekeeping, testing proceeds if at least one hypothesis in $F_k$ is rejected. Within families, standard methods (e.g., Holm or Hochberg) can be applied to control the intra-family FWER.

More generally, gatekeeping strategies can be combined with alpha-propagation or fallback principles, allowing unspent significance levels to flow between families under pre-specified logical rules (Bretz et al., 2009). These designs preserve the strong control of the FWER under arbitrary dependence among test statistics, assuming that the structure of the hypothesis families and transitions is fully specified before data analysis.

## 4. Empirical Analysis: Housing, Health, and Happiness

Cattaneo et al. (2009) evaluate the Piso Firme program in Mexico, which aimed to replace dirt floors with cement floors in low-income households. The study investigates the program's effects on child health, cognitive development, and adult welfare, using a quasi-experimental design that compares treated households to a control group in geographically similar areas, while adjusting for differences in pretreatment characteristics.

The authors examine the impact of Piso Firme across multiple outcomes, including child health indicators and measures of adult well-being. As a first step, they document that the program significantly increased the prevalence of cement floors in treated households—a necessary precondition for the program to affect downstream health and welfare outcomes. The results show that replacing dirt floors with cement substantially reduced parasitic infestations, diarrhea, and anemia among children, while improving cognitive development. Among adults, the program improved mental health, reduced stress, and greater satisfaction with housing and overall quality of life. These findings underscore the role of housing improvements as an effective and low-cost policy tool for enhancing public health and well-being.

The analysis is structured around the sequential effects of Piso Firme: first, on parasitic infections, diarrhea, and anemia; and subsequently on child cognitive development and adult mental health (depression and stress). The authors estimate these effects using multiple model specifications: Model 1 includes no covariates; Model 2 adds demographic and health controls, while Model 3 further includes controls for participation in other public social programs. The results are broadly consistent across models; thus, we present only those from Model 3 here, with the full set of estimates available in the appendix.

## 4.1 Main results

Table 1 replicates Table 4 in Cattaneo et al. (2009), examining the program's impact on the adoption of cement floors. Outcomes include the share of rooms with cement floors, and indicators for whether specific rooms (kitchen, dining room, bathroom, bedroom) have cement flooring. The results in Columns 1-4 show that Piso Firme induced floor replacement across treated households, with highly robust estimates across all model specifications (Table 1A, in the appendix). Columns 1-4 in Table 2 (Table 5 in Cattaneo et al., 2009) reports intention-to-treat estimates of the program's effects on child health outcomes for children younger than six years old. The outcomes analyzed include parasitic infections, diarrhea, anemia, height, weight, and cognitive development. The program led to significant reductions in parasites, diarrhea, anemia, and improvements in cognitive development. However, there were no significant effects on standard height and weight anthropometric measures.

Further analyses show that these health findings are consistent with the hypothesized mechanism: replacing dirt floors with cement disrupts the transmission of parasites, which in turn reduces diarrhea and anemia, with anemia reductions contributing to cognitive gains.

***Conventional Multiple Testing Corrections***

Tables 1 and 2 (Columns 5a-d) also report the original and adjusted p-values that control the FWER using some of the most common methods implemented in empirical practice: the free step-down resampling method of Westfall and Young (1993), and the Bonferroni-Holm and Sidak-Holm adjusted p-values. In Table 1, all estimates remain highly statistically significant under all correction methods. In contrast, the results in Table 2 are more sensitive to multiplicity adjustments: the effects on parasite count and diarrhea lose significance even at the 10% level, and the evidence for cognitive gains becomes weaker and mixed across specifications (Table 2A).

A limitation of these conventional multiple testing corrections is that they do not account for the hierarchical structure of the hypotheses. To address this, in the next section we implement sequential procedures that reflect the underlying causal sequence of outcomes. Figure 1 illustrates the hypothesized causal pathway: Piso Firme increases the probability of cement floor installation, which reduces exposure to parasites and associated health conditions such as diarrhea and anemia. These health improvements enhance child cognitive development and contribute to better parental mental health, ultimately improving overall household welfare.

***Hierarchical Multiple Testing Corrections***

We first implement a fallback procedure as described in Section 3, which controls the FWER while accounting for the hierarchical structure of our hypotheses. Specifically, we test the primary hypothesis—that Piso Firme significantly increased cement floor coverage—before proceeding to secondary hypotheses concerning child health outcomes. Results, summarized in Table 3, show strong statistical evidence supporting Piso Firme's effect on cement floor coverage, satisfying the fallback procedure's initial gate. This confirmation of the primary outcome permits subsequent testing hypotheses with greater power, supporting the original findings (at the 10% significance level) related to improvements in children's health indicators, such as reductions in parasitic infections and diarrhea prevalence. This hierarchical approach aligns directly with the program's causal logic that observed health benefits are plausibly mediated by the program's success in improving housing conditions.

To complement our analysis, we implement a sequential gatekeeping procedure for the Piso Firme evaluation, structured around two ordered families of hypotheses. Family 1 (Fk1) includes housing outcomes, specifically measures of cement floor coverage, Family 2 (Fk2) encompasses child health outcomes, including parasitic infections, diarrhea prevalence, and anemia rates, and Family 3 (Fk3) includes cognitive development measures. In this sequential gatekeeping framework, we first test hypotheses in Fk1, applying multiple testing corrections (e.g., Holm's method) to control the FWER within the housing outcomes.

Testing proceeds to Family 2 only if at least one hypothesis in Fk1 is rejected at the predefined significance level; and we proceed similarly with Fk3. This design reflects the hypothesized causal pathway from improved housing conditions to better child health outcomes and ensures that inference on secondary outcomes (Fk2, Fk3) is contingent on confirming the intervention's primary effect on housing. By structuring the hypothesis testing hierarchically in this way, we control type I errors while aligning the statistical analysis with the logical progression of expected impacts in the Piso Firme program.

This sequential gatekeeping approach is well suited to the Piso Firme evaluation because it aligns directly with the program's causal logic: the intervention's primary goal is to improve housing quality by increasing cement floor coverage (Family 1), which is hypothesized to reduce exposure to disease vectors and environmental contaminants. Only if Piso Firme demonstrably achieves this foundational housing improvement does it make sense to evaluate downstream effects on child health outcomes (Family 2), such as reductions in parasitic infections, diarrhea, and anemia; and further down on cognitive development (Family 3). Structuring the testing hierarchically in this way prevents inflated type I error rates from testing many outcomes indiscriminately, while focusing statistical power on detecting meaningful effects along the hypothesized causal pathway from housing improvements to health impacts.

The results from this analysis are presented in Table 4. Using a sequential gatekeeping procedure with both Holm's and Westfall–Young's corrections, we find strong evidence (at the 10% significance level) that the Piso Firme program had a positive impact across all three families of outcomes. Cement floor coverage (Family 1) was broadly significant, opening the gate to health outcomes (Family 2), where all three measures—anemia, diarrhea, and parasite count—remained statistically significant after adjustment. This, in turn, justified testing cognitive outcomes (Family 3), where both measures—the MacArthur and Peabody vocabulary scores—were also statistically significant. These results support a full causal chain from infrastructure improvements to child health and early development. This reinforces the robustness of the findings under more sophisticated multiple testing adjustments.

## 5. Other applications

Hierarchical structures among hypotheses are very common in both the social and medical sciences, as they help clarify the pathways through which interventions affect outcomes. Here we highlight their role by briefly discussing a few other examples to illustrate how hierarchical testing provides insights into mechanisms and ensures more rigorous evaluations.

Galiani et al. (2011) study how mandatory military conscription in Argentina causally affects subsequent criminal behavior, using random lottery assignment as a natural experiment. They find conscription significantly increases the probability of developing a criminal record, especially for property and white-collar crimes. The authors also document negative impacts on future labor market outcomes, such as lower formal employment and reduced earnings. The analysis employs a structured framework with clear causal pathways across multiple hypotheses: draft eligibility directly increases conscription likelihood, conscription causally leads to higher crime rates, and conscription negatively impacts socioeconomic outcomes, including labor market participation and earnings. Draft eligibility alone, without conscription, shows limited direct impacts on socioeconomic outcomes.

Duflo et al. (2012) examine whether tying teacher salaries to attendance through a monitoring and incentive program reduces absenteeism and improves student performance in India, using a randomized control trial. The study tests a clear causal pathway: improved teacher attendance (H1) should lead to better student outcomes (H2). For outcomes measured at mid-term, H1 assesses the program's immediate effect on reducing absenteeism, while H2 evaluates whether this translated into higher student test scores. For outcomes after the post-term test, H1 examines whether reductions in absenteeism persisted, and H2 tests if improved attendance sustained gains in student achievement.

Galiani et al. (2015) evaluate a large-scale handwashing intervention in Peru, which included community-based caregiver education and school-based handwashing curricula. They find the intervention significantly increased knowledge and improved the determinants influencing handwashing behaviors. These intermediate outcomes led to measurable improvements in actual handwashing practices, particularly the critical behavior of washing hands with soap and water before eating. However, despite successfully changing behaviors, the program did not show statistically significant improvements in child health outcomes, as measured by parasite prevalence, suggesting a breakdown in the causal pathway from improved hygiene behaviors to health improvements.[2]

Carpio et al. (2025) investigate how social health insurance (SHI) affects student academic performance in Peru using a sharp Regression Discontinuity Design (RDD) based on welfare eligibility. They find that SHI significantly improves standardized test scores by approximately 0.96 standard deviations in reading and 0.84 in mathematics. These educational gains primarily result from improved child and parental health—particularly reduced anemia—resulting from increased healthcare utilization, including doctor visits, medications, dental care, vaccinations, and iron supplementation. Although healthier parents have more productive time, this does not translate into higher household income or educational spending. The authors propose a hierarchical framework (Figure 2) to illustrate the causal pathways, highlighting that in contexts with initially low healthcare utilization, SHI enhances educational outcomes mainly through improved child health, secondarily through better parental health, and least importantly through financial protection, which may have ambiguous effects due to induced demand.

Galiani et al. (2017) provide empirical evidence regarding the causal effects that upgrading slum dwellings has on the living conditions of the extremely poor. They study the impact of providing better houses *in situ* to slum dwellers in El Salvador, Mexico and Uruguay. They experimentally evaluate the impact of a housing project run by the NGO TECHO ("roof"), which supplies basic prefabricated houses to members of extremely poor population groups in Latin America. The main objective of the program is to improve household well-being. Their findings show that better houses have a positive effect on overall housing conditions and general well-being: the members of treated households are happier with their quality of life. In two countries, they also document improvements in children's health; in El Salvador, slum dwellers who have received the TECHO houses also feel that they are safer. They do not find this result, however, in the other two experimental samples. There are no other noticeable robust effects in relation to the possession of durable goods or labor outcomes. Notably, their results are robust in terms of both their internal and external validity because they are derived from similar experiments in three different Latin American countries.[3]

[2] In line with the approach in our paper, the authors follow Kling et al. (2007) in constructing summary indices by family of outcome variables.

Furthermore, we explore these issues through the analysis in Young (2019), who systematically reexamines a large number of experimental studies using randomization inference and Bonferroni-type corrections. Table 5 summarizes a selection of these studies. His approach assumes that all reported outcomes are independent and equally important for inference, without considering the logical or causal structure that often links them. As a result, his corrections are deliberately conservative and frequently render statistically insignificant findings that were originally reported as significant. While the methodology offers rigorous control of the FWER, it imposes a uniform adjustment that treats all hypotheses as exchangeable. This overlooks the fact that many of the outcomes in these evaluations are causally ordered. In such cases, hierarchical procedures offer a more appropriate framework by reflecting the structure of the underlying theory or program design and by concentrating inferential adjustments on the relevant segments of the causal chain.

## 6. Conclusion

Multiple testing corrections are essential for controlling false discoveries in empirical research, but classical methods such as Bonferroni or Holm adjustments often sacrifice statistical power by treating all hypotheses as independent and equally important. Hierarchical approaches offer a more nuanced framework by explicitly incorporating logical or causal relationships among hypotheses, allowing researchers to prioritize tests according to their position within a theoretical or causal model. This structure enables more powerful and interpretable inference by concentrating adjustments on meaningful hypothesis families or sequences. Hierarchical methods are particularly valuable in applications with structured testing problems, such as clinical trials, where effects on biomarkers or intermediate outcomes precede final endpoints, or program evaluations, where interventions may first influence behaviors before affecting long-term outcomes. By respecting these relationships, hierarchical multiple testing helps researchers draw clearer conclusions about where interventions succeed or fail along the causal pathway, while maintaining rigorous control of type I error.

---

[3] In the spirit of our paper, this paper applies Bonferroni corrections to control the FWER, adjusting p-values based on the number of outcomes within conceptually related blocks, grouped by table and by country experiment. The authors also follow Kling et al. (2007) in constructing summary indices by outcome family.

Figure 1. Piso Firme: Causal Pathways

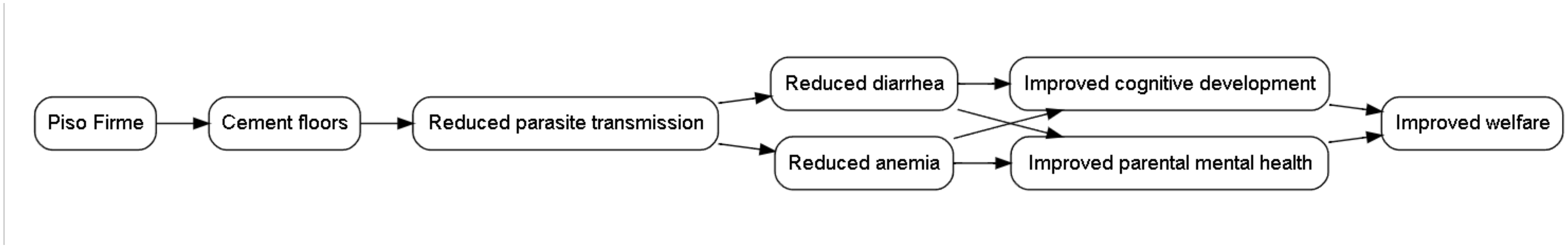

*Notes:* Cattaneo et al. (2009)

Figure 2: Causal diagram of the effect of social health insurance on student performance from Carpio, Gomez, and Lavado (2025).

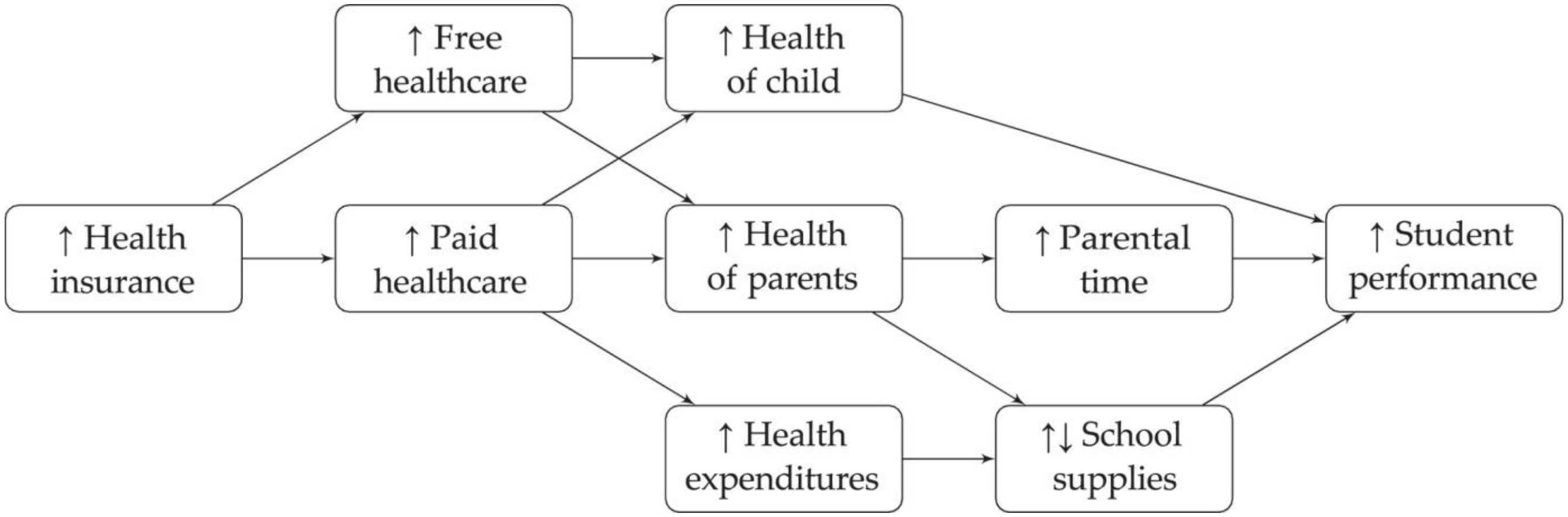

*Notes:* The graph outlines the flow of the causal associations, starting from SHI coverage and branching to various consumption variables, health and financial status, educational inputs, and student performance. Prepared by the authors.

Table 1. Piso Firme: Regressions of Cement Floor Coverage Measures on Program Dummy

| (1) Dependent Variable | (2) Control Group Mean (SD) | (3) Coef. | (4) SE [Signif] | (5) p-value | | | |
|---|---|---|---|---|---|---|---|
| | | | | (a) Original | (b) Westfall-Young | (c) Bonferroni-Holm | (d) Sidak-Holm |
| Share of rooms with cement floors | 0.728 (0.363) | 0.21 | [0.019]*** | 0.000 | 0.00 | 0.00 | 0.00 |
| Cement floor in kitchen | 0.671 (0.470) | 0.265 | [0.023]*** | 0.000 | 0.00 | 0.00 | 0.00 |
| Cement floor in dining room | 0.709 (0.455) | 0.221 | [0.025]*** | 0.000 | 0.00 | 0.00 | 0.00 |
| Cement floor in bathroom | 0.803 (0.398) | 0.117 | [0.018]*** | 0.000 | 0.00 | 0.00 | 0.00 |
| Cement floor in bedroom | 0.668 (0.471) | 0.245 | [0.020]*** | 0.000 | 0.00 | 0.00 | 0.00 |

*Notes:* Cattaneo et al. (2009)

Table 2. Piso Firme: Regressions of Children's Health Measures on Program Dummy

| (1) Dependent Variable | (2) Control Group Mean (SD) | (3) Coef. | (4) SE [Signif] | (5) p-value | | | |
|---|---|---|---|---|---|---|---|
| | | | | (a) Original | (b) Westfall-Young | (c) Bonferroni-Holm | (d) Sidak-Holm |
| Parasite count | 0.333 (0.673) | -0.06 | [0.032]** | 0.046 | 0.195 | 0.186 | 0.173 |
| Diarrhea | 0.142 (0.349) | -0.02 | [0.009]* | 0.046 | 0.195 | 0.186 | 0.173 |
| Anemia | 0.426 (0.495) | -0.08 | [0.027]*** | 0.002 | 0.016 | 0.015 | 0.015 |
| MacArthur Communicative Development Test score | 13.354 (18.952) | 5.557 | [1.641]*** | 0.001 | 0.008 | 0.007 | 0.007 |
| Picture Peabody Vocabulary Test percentile score | 30.656 (24.864) | 3.083 | [1.410]** | 0.029 | 0.148 | 0.153 | 0.144 |
| Height-for-age z-score | -0.605 (1.104) | 0.002 | [0.039] | 0.959 | 0.959 | 1.000 | 0.960 |
| Weight-for-height z-score | 0.125 (1.133) | -0.01 | [0.037] | 0.766 | 0.953 | 1.000 | 0.945 |

*Notes:* Cattaneo et al. (2009). Highlighted cells: significance at 10% level.

Table 3. Piso Firme: Fallback procedure to adjust for multiple testing

| (1)<br>Dependent Variable | (2)<br>Unadjusted p-value | (3)<br>Weights | (4)<br>Initial alpha | (4)<br>alpha propagation | (5)<br>H0 |
|---|---|---|---|---|---|
| **Cement Floor Coverage Measures** | | | | | |
| Share of rooms with cement floors | 0.000 | 0.100 | 0.010 | 0.010 | Reject |
| Cement floor in kitchen | 0.000 | 0.100 | 0.010 | 0.020 | Reject |
| Cement floor in dining room | 0.000 | 0.100 | 0.010 | 0.030 | Reject |
| Cement floor in bathroom | 0.000 | 0.100 | 0.010 | 0.040 | Reject |
| Cement floor in bedroom | 0.000 | 0.100 | 0.010 | 0.050 | Reject |
| **Children's Health Measures** | | | | | |
| Parasite count | 0.046 | 0.100 | 0.010 | 0.060 | Reject |
| Diarrhea | 0.046 | 0.100 | 0.010 | 0.070 | Reject |
| Anemia | 0.002 | 0.100 | 0.010 | 0.080 | Reject |
| MacArthur Communicative Development Test score | 0.001 | 0.100 | 0.010 | 0.090 | Reject |
| Picture Peabody Vocabulary Test percentile score | 0.029 | 0.100 | 0.010 | 0.100 | Reject |
| Height-for-age z-score | 0.959 | 0.000 | 0.000 | 0.100 | No Reject |
| Weight-for-height z-score | 0.766 | 0.000 | 0.000 | 0.000 | No Reject |

*Notes:* Cattaneo et al. (2009). Highlighted cells: significance at 10% level.

Table 4. Piso Firme: Gatekeeping procedures to adjust for multiple testing

| (1) Family | (2) Dependent Variable | (3) p-value | | | |
|---|---|---|---|---|---|
| | | (a) Original | (b) Westfall-Young | (c) Bonferroni-Holm | (d) Sidak-Holm |
| **$F_1$** | Cement floor coverage (various measures) | 0.000 | 0.000 | 0.000 | 0.000 |
| | *(Gate to $F_2$ opened)* | | | | |
| **$F_2$** | Anemia | 0.002 | 0.090 | 0.093 | 0.091 |
| | Diarrhea | 0.046 | 0.090 | 0.093 | 0.091 |
| | Parasite count | 0.046 | 0.009 | 0.007 | 0.007 |
| | *(Gate to $F_3$ opened)* | | | | |
| **$F_3$** | MacArthur Communicative Development Score | 0.010 | 0.003 | 0.002 | 0.002 |
| | Peabody Vocabulary Test Percentile Score | 0.029 | 0.032 | 0.031 | 0.031 |

*Notes:* Cattaneo et al. (2009). Highlighted cells: significance at 10% level.

Table 5: Overview of recent papers

| | Paper | Causal Relationships |
|---|---|---|
| 1 | Angrist, Joshua, Daniel Lang, and Philip Oreopoulos. 2009. “Incentives and Services for College Achievement: Evidence from a Randomized Trial.” American Economic Journal: Applied Economics 1 (1): 136–163. | Offering scholarships (incentives) increases the likelihood that students will use academic services.<br>Women who receive both services and incentives show significant improvement in their grades, credits earned, and academic probation rates compared to those who receive neither intervention. |
| 2 | Angrist, Joshua, and Victor Lavy. 2009. “The Effects of High Stakes High School Achievement Awards: Evidence from a Randomized Trial.” American Economic Review 99 (4): 1384–1414. | Incentives increased certification, and subsequent postsecondary enrollment. |
| 3 | Burde, Dana and Leigh L. Linden. 2013. “Bringing Education to Afghan Girls: A Randomized Controlled Trial of Village-Based Schools.” American Economic Journal: Applied Economics 5 (3): 27–40. | Placing a school in the village ➔ Increased enrollment (both girls and boys).<br>Increased enrollment ➔ Improved academic performance (test scores). |
| 4 | Cai, Hongbin, Yuyu Chen, and Hanming Fang. 2009. “Observational Learning: Evidence from a Randomized Natural Field Experiment.” American Economic Review 99 (3): 864–882. | Impact of showing the top 5 most popular dishes on demand for those dishes -> impact on dinner experience |
| 5 | Carpenter, Jeffrey, Peter Hans Matthews, and John Schirm. 2010. “Tournaments and Office Politics: Evidence from a Real Effort Experiment.” American Economic Review 100 (1): 504–517. | How workers respond to different incentive structures (piece rates vs. tournaments) in terms of effort levels.<br>The experiment assessed whether output quality changes depending on the incentive structure, considering the influence of sabotage as a potential mediator.<br>Also investigated whether the presence of tournaments increases the likelihood of sabotage or office politics among workers, thereby influencing their effort and the quality of their output |
| 6 | Duflo, Esther, Rema Hanna, and Stephen P. Ryan. 2012. “Incentives Work: Getting Teachers to Come to School.” American Economic Review 102 (4): 1241–1278. | test whether monitoring and financial incentives can reduce teacher absence and increase learning in India |
| 7 | Dupas, Pascaline and Jonathan Robinson. 2013. “Why Don’t the Poor Save More? Evidence from Health Savings Experiments.” American Economic Review 103 (4): 1138– 1171. | Impact of simple saving (e.g safe box) technologies on investment in preventative health.<br>The introduction of safe boxes influences participation in informal saving networks. |
| 8 | Fehr, Ernst and Lorenze Goette. 2007. "Do Workers Work More if Wages Are High? Evidence from a Randomized Field Experiment." American Economic Review 97 (1): 298- 317 | Wage increase -> increase in working hours<br>Wage increase -> decrease in daily effort<br>Combined effect of increased working hours and decreased effort -> net increase in labor supply |
| 9 | Giné, Xavier, Jessica Goldberg, and Dean Yang. 2012. "Credit Market Consequences of Improved Personal Identification: Field Experimental Evidence from Malawi." American Economic Review 102 (6): 2923–2954. | Credit market impacts of improved personal identification through fingerprinting in rural Malawi<br>Impact of fingerprints on repayment<br>Impact of fingerprinting on agricultural inputs and profits |
| 10 | Macours, Karen, Norbert Schady, and Renos Vakis. 2012. "Cash Transfers, Behavioral Changes, and Cognitive Development in Early Childhood: Evidence from a Randomized Experiment." American Economic Journal: Applied Economics 4 (2): 247–273. | Impact of cash transfers on cognitive development in early childhood: food, stimulation (impact on early enrollment), preventive health |
| 11 | de Mel, Suresh, David McKenzie, and Christopher Woodruff. 2013. "The Demand for, and Consequences of, Formalization among Informal Firms in Sri Lanka." American Economic Journal: Applied Economics 5 (2): 122–150. | Impact of incentives to formalize on informal firms<br>Impact of formalizing on profits. |
| 12 | Oster, Emily and Rebecca Thornton. 2011. "Menstruation, Sanitary Products, and School Attendance: Evidence from a Randomized Evaluation." American Economic Journal: Applied Economics 3 (1): 91–100. | Impact of menstruation on school attendance<br>Improving sanitary technology on reducing girls' attendance |
| 13 | Thornton, Rebecca L. 2008. "The Demand for, and Impact of, Learning HIV Status." American Economic Review 98 (5): 1829–1863. | Effects of monetary incentives to learn HIV results after being tested<br>Effects of learning about HIV on purchasing condoms |
| | | |
| | | |
| | | |

# Appendix Table 1A

| Dependent Variable | Control Group Mean (SD) | Model 1 | | | | | | Model 2 | | | | | | Model 3 | | | | | |
|---|---|---|---|---|---|---|---|---|---|---|---|---|---|---|---|---|---|---|---|
| | | Coef. | SE [Signif] | p-value | | | | Coef. | SE [Signif] | p-value | | | | Coef. | SE [Signif] | p-value | | | |
| | | | | Unadjusted | Westfall-Young | Bonferroni-Holm | Sidak-Holm | | | Original | Westfall-Young | Bonferroni-Holm | Sidak-Holm | | | Original | Westfall-Young | Bonferroni-Holm | Sidak-Holm |
| Share of rooms with cement floors | 0.728 (0.363) | 0.202 | [0.021] *** | 0.000 | 0.00 | 0.00 | 0.00 | 0.208 | [0.019] *** | 0.000 | 0.00 | 0.00 | 0.00 | 0.21 | [0.019] *** | 0.000 | 0.00 | 0.00 | 0.00 |
| Cement floor in kitchen | 0.671 (0.470) | 0.255 | [0.025] *** | 0.000 | 0.00 | 0.00 | 0.00 | 0.26 | [0.023] *** | 0.000 | 0.00 | 0.00 | 0.00 | 0.265 | [0.023] *** | 0.000 | 0.00 | 0.00 | 0.00 |
| Cement floor in dining room | 0.709 (0.455) | 0.21 | [0.026] *** | 0.000 | 0.00 | 0.00 | 0.00 | 0.217 | [0.025] *** | 0.000 | 0.00 | 0.00 | 0.00 | 0.221 | [0.025] *** | 0.000 | 0.00 | 0.00 | 0.00 |
| Cement floor in bathroom | 0.803 (0.398) | 0.105 | [0.022] *** | 0.000 | 0.00 | 0.00 | 0.00 | 0.113 | [0.018] *** | 0.000 | 0.00 | 0.00 | 0.00 | 0.117 | [0.018] *** | 0.000 | 0.00 | 0.00 | 0.00 |
| Cement floor in bedroom | 0.668 (0.471) | 0.238 | [0.020] *** | 0.000 | 0.00 | 0.00 | 0.00 | 0.245 | [0.021] *** | 0.000 | 0.00 | 0.00 | 0.00 | 0.245 | [0.020] *** | 0.000 | 0.00 | 0.00 | 0.00 |

# Appendix Table 2A

| | | Model 1 | | | | | | Model 2 | | | | | | Model 3 | | | | | |
|---|---|---|---|---|---|---|---|---|---|---|---|---|---|---|---|---|---|---|---|
| **Dependent Variable** | **Control Group Mean (SD)** | **Coef.** | **SE [Signif]** | **p-value** | | | | **Coef.** | **SE [Signif]** | **p-value** | | | | **Coef.** | **SE [Signif]** | **p-value** | | | |
| | | | | **Unadjusted** | **Westfall-Young** | **Bonferroni-Holm** | **Sidak-Holm** | | | **Unadjusted** | **Westfall-Young** | **Bonferroni-Holm** | **Sidak-Holm** | | | **Unadjusted** | **Westfall-Young** | **Bonferroni-Holm** | **Sidak-Holm** |
| Parasite count | 0.333 (0.673) | -0.065 | [0.032]** | 0.042 | 0.203 | 0.223 | 0.204 | -0.064 | [0.031]** | 0.039 | 0.131 | 0.134 | 0.127 | -0.064 | [0.032]** | 0.046 | 0.195 | 0.186 | 0.173 |
| Diarrhea | 0.142 (0.349) | -0.018 | [0.009]* | 0.046 | 0.215 | 0.223 | 0.204 | -0.02 | [0.009]** | 0.026 | 0.131 | 0.134 | 0.127 | -0.018 | [0.009]* | 0.046 | 0.195 | 0.186 | 0.173 |
| Anemia | 0.426 (0.495) | -0.085 | [0.028]*** | 0.002 | 0.025 | 0.020 | 0.020 | -0.081 | [0.027]*** | 0.003 | 0.019 | 0.019 | 0.019 | -0.083 | [0.027]*** | 0.002 | 0.016 | 0.015 | 0.015 |
| MacArthur Communicative Development Test score | 13.354 (18.952) | 4.031 | [1.650]** | 0.015 | 0.096 | 0.097 | 0.093 | 5.652 | [1.642]*** | 0.001 | 0.007 | 0.006 | 0.006 | 5.557 | [1.641]*** | 0.001 | 0.008 | 0.007 | 0.007 |
| Picture Peabody Vocabulary Test percentile score | 30.656 (24.864) | 2.668 | [1.689] | 0.114 | 0.317 | 0.350 | 0.311 | 3.206 | [1.430]** | 0.025 | 0.122 | 0.134 | 0.127 | 3.083 | [1.410]** | 0.029 | 0.148 | 0.153 | 0.144 |
| Height-for-age z-score | -0.605 (1.104) | 0.007 | [0.043] | 0.871 | 0.986 | 1.000 | 0.983 | -0.002 | [0.038] | 0.958 | 0.984 | 1.000 | 0.987 | 0.002 | [0.039] | 0.959 | 0.959 | 1.000 | 0.960 |
| Weight-for-height z-score | 0.125 (1.133) | 0.002 | [0.034] | 0.953 | 0.986 | 1.000 | 0.983 | -0.005 | [0.036] | 0.890 | 0.984 | 1.000 | 0.987 | -0.011 | [0.037] | 0.766 | 0.953 | 1.000 | 0.945 |

# Appendix Table 3A

| Dependent Variable | Model 1 Unadjusted p-value | Model 1 weights | Model 1 Initial alpha | Model 1 alpha propagation | Model 1 H0 | Model 2 Unadjusted p-value | Model 2 weights | Model 2 Initial alpha | Model 2 alpha propagation | Model 2 H0 | Model 3 Unadjusted p-value | Model 3 weights | Model 3 Initial alpha | Model 3 alpha propagation | Model 3 H0 |
|---|---|---|---|---|---|---|---|---|---|---|---|---|---|---|---|
| Share of rooms with cement floors | 0.000 | 0.100 | 0.010 | 0.010 | Reject | 0.000 | 0.100 | 0.010 | 0.010 | Reject | 0.000 | 0.100 | 0.010 | 0.010 | Reject |
| Cement floor in kitchen | 0.000 | 0.100 | 0.010 | 0.020 | Reject | 0.000 | 0.100 | 0.010 | 0.020 | Reject | 0.000 | 0.100 | 0.010 | 0.020 | Reject |
| Cement floor in dining room | 0.000 | 0.100 | 0.010 | 0.030 | Reject | 0.000 | 0.100 | 0.010 | 0.030 | Reject | 0.000 | 0.100 | 0.010 | 0.030 | Reject |
| Cement floor in bathroom | 0.000 | 0.100 | 0.010 | 0.040 | Reject | 0.000 | 0.100 | 0.010 | 0.040 | Reject | 0.000 | 0.100 | 0.010 | 0.040 | Reject |
| Cement floor in bedroom | 0.000 | 0.100 | 0.010 | 0.050 | Reject | 0.000 | 0.100 | 0.010 | 0.050 | Reject | 0.000 | 0.100 | 0.010 | 0.050 | Reject |
| Parasite count | 0.042 | 0.100 | 0.010 | 0.060 | Reject | 0.039 | 0.100 | 0.010 | 0.060 | Reject | 0.046 | 0.100 | 0.010 | 0.060 | Reject |
| Diarrhea | 0.046 | 0.100 | 0.010 | 0.070 | Reject | 0.026 | 0.100 | 0.010 | 0.070 | Reject | 0.046 | 0.100 | 0.010 | 0.070 | Reject |
| Anemia | 0.002 | 0.100 | 0.010 | 0.080 | Reject | 0.003 | 0.100 | 0.010 | 0.080 | Reject | 0.002 | 0.100 | 0.010 | 0.080 | Reject |
| MacArthur Communicative Development Test score | 0.015 | 0.100 | 0.010 | 0.090 | Reject | 0.001 | 0.100 | 0.010 | 0.090 | Reject | 0.001 | 0.100 | 0.010 | 0.090 | Reject |
| Picture Peabody Vocabulary Test percentile score | 0.114 | 0.100 | 0.010 | 0.100 | No Reject | 0.025 | 0.100 | 0.010 | 0.100 | Reject | 0.029 | 0.100 | 0.010 | 0.100 | Reject |
| Height-for-age z-score | 0.871 | 0.000 | 0.000 | 0.000 | No Reject | 0.958 | 0.000 | 0.000 | 0.100 | No Reject | 0.959 | 0.000 | 0.000 | 0.100 | No Reject |
| Weight-for-height z-score | 0.953 | 0.000 | 0.000 | 0.000 | No Reject | 0.890 | 0.000 | 0.000 | 0.000 | No Reject | 0.766 | 0.000 | 0.000 | 0.000 | No Reject |